\documentclass[twoside]{articlek}

\usepackage[normal]{caption}

\renewcommand{\refname}{REFERENCES}

\def\BibTeX{{\rm B\kern-.05em{\sc i\kern-.025em b}\kern-.08em
		T\kern-.1667em\lower.7ex\hbox{E}\kern-.125emX}}

\usepackage{graphicx}

\renewcommand{\refname}{REFERENCES}

\def\BibTeX{{\rm B\kern-.05em{\sc i\kern-.025em b}\kern-.08em
		T\kern-.1667em\lower.7ex\hbox{E}\kern-.125emX}}

\usepackage{graphicx}

\begin{document}
%
%

\begin{flushleft}
	{\large\bf Some remarks on the discovery of $^{244}$Md
		}
	\vspace*{25pt}

	{\bf Fritz Peter He\ss berger$^{1,2,}\footnote{E-mail: \texttt{f.p.hessberger@gsi.de}}$} \\
\vspace{2pt}
{$^1$GSI - Helmholtzzentrum f\"ur Schwerionenforschung GmbH, Planckstra\ss e 1, 64291 Darmstadt, Germany\\
	$^2$Helmholtz Institut Mainz, Johann-Joachim-Becherweg, 55128 Mainz, Germany}\\
\vspace{5pt}
{\bf Michael Block$^{1,2,3}$} \\
\vspace{2pt}
{$^1$GSI - Helmholtzzentrum f\"ur Schwerionenforschung GmbH, Planckstra\ss e 1, 64291 Darmstadt, Germany\\
	$^2$Johannes Gutenberg-Universit\"at Mainz, 55099 Mainz, Germany \\
$^3$Helmholtz Institut Mainz, Johann-Joachim-Becherweg, 55128 Mainz, Germany}\\
\vspace{5pt}
{\bf Christoph D\"ullmann$^{1,2,3}$} \\
\vspace{2pt}
{$^1$Johannes Gutenberg-Universit\"at Mainz, 55099 Mainz, Germany \\
	$^2$GSI - Helmholtzzentrum f\"ur Schwerionenforschung GmbH, Planckstra\ss e 1, 64291 Darmstadt, Germany\\
	$^3$Helmholtz Institut Mainz, Johann-Joachim-Becherweg, 55128 Mainz, Germany}\\
\vspace{5pt}
{\bf Alexander Yakushev$^{1}$} \\
\vspace{2pt}
{$^1$GSI - Helmholtzzentrum f\"ur Schwerionenforschung GmbH, Planckstra\ss e 1, 64291 Darmstadt, Germany}\\
\vspace{5pt}
{\bf Matti Leino$^{1}$}\\
\vspace{2pt}
{$^1$University Jyv\"askyl\"a, 40014 Jyv\"askyl\"a, Finland} \\
\vspace{5pt}
{\bf Juha Uusitalo$^{1}$}\\
\vspace{2pt}
{$^1$University Jyv\"askyl\"a, 40014 Jyv\"askyl\"a, Finland} \\

\end{flushleft}
          
Version: December 18, 2020\\

	\begin{abstract}
		
		In two recent papers by Pore et al.\cite{Pore20} and Khuyagbaatar et al.\cite{Khuyag20} discovery of the new isotope $^{244}$Md was reported. The decay data, however,
		are conflicting. While Pore et al. \cite{Pore20} report two isomeric states decaying by $\alpha$ emission with E$_{\alpha}$(1)\,=\,8.66(2) MeV, 
		T$_{1/2}$(1)\,=\,0.4$^{+0.4}_{-0.1}$s and E$_{\alpha}$(2)\,=\,8.31(2) MeV, 
		T$_{1/2}$(2)$\approx$6 s,  Khuyagbaatar et al. \cite{Khuyag20} report only a single transition with a broad energy distribution 
		of E$_{\alpha}$\,=\,(8.73\,-\,8.86) MeV and T$_{1/2}$\,=\,0.30$^{+0.19}_{-0.09}$ s. The data published in \cite{Pore20} are
		very similar to those published for $^{245m}$Md (E$_{\alpha}$\,=\,8.64(2), 8.68(2) MeV, T$_{1/2}$\,=\,0.35$^{+0.23}_{-0.16}$ s \cite{Ninov96}). Therefore, we compare the
		data presented for $^{244}$Md in \cite{Pore20}  with those reported for $^{245}$Md in \cite{Ninov96} and also in \cite{Khuyag20}. 
		We conclude that the data presented in \cite{Pore20} shall be attributed to $^{245}$Md with small contributions (one event each)
		from $^{245}$Fm and probably $^{246}$Md. 
		
	\end{abstract}


	
	\section{Introduction}
	\label{intro}
	Discovery of $^{244}$Md was first reported by J.L. Pore et al. \cite{Pore20}. They used the reaction $^{209}$Bi($^{40}$Ar,5n)$^{244}$Md at a bombarding energy
	of $\approx$220 MeV, which corresponds to an excitation energy of the compound nucleus $^{249}$Md of E$^{*}$\,$\approx$\,46 MeV at a production in the center 
	of the target. 
	They observed four events after the mass spectrometer FIONA at a position where events with mass number A\,=\,244 were expected, and six $\alpha$ decay chains
	in the BGS focal plane detector. The latter were attributed to the decay of two states in 
	$^{244}$Md, one with E$_{\alpha}$\,=\,8.308$\pm$0.019 MeV, T$_{1/2}$\,$\approx$6 s (1 event), and one 
	with E$_{\alpha}$\,=\,8.663$\pm$0.023 MeV, T$_{1/2}$\,=\,0.4$^{+0.4}_{-0.1}$ s (4 events).\\
	In a publication by J. Khuyagbaatar et al. identification of $^{244}$Md was reported using the reaction $^{197}$Au($^{50}$Ti, 3n)$^{244}$Md \cite{Khuyag20}.
	The experiment was peformed at two bombarding energies of 239.8 MeV and 231.5 MeV (center of target), corresponding to excitation energies of E$^{*}$\,=\,32.7 MeV 
	and E$^{*}$\,=\,26.2 MeV. They reported two $\alpha$ acitivities. One, with an energy range E$_{\alpha}$\,=\,(8.7\,-\,8.8) MeV and a half-life of T$_{1/2}$\,=\,
	0.30$^{+0.19}_{-0.09}$ s was observed only at the higher excitation energy (7 events); the second activity was observed at both energies (three events each
	with full energy release in the stop detector) within
	an energy range of E$_{\alpha}$\,=\,(8.6\,-\,8.7) MeV and a half-life of T$_{1/2}$\,=\,0.33$^{+0.15}_{-0.08}$ s. This activity was attributed to the previously
	reported isotope $^{245}$Md.\\
	The isotope $^{245}$Md was first observed in an experiment performed at the velocity filter SHIP at GSI, Darmstadt, Germany, using the reaction $^{209}$Bi($^{40}$Ar,4n)$^{245}$Md at a bombarding energy of 5.12 AMeV
	(204.8 MeV) corresponding to an excitation energy of E$^{*}$\,=\,40 MeV \cite{Ninov96}. The authors reported two $\alpha$ energies of E$_{\alpha}$\,=\,8640$\pm$20, 
	8680\,$\pm$20 keV, and a half-life of T$_{1/2}$\,=\,0.35$^{+0.23}_{-0.18}$ s and also a spontaneous fission activity of T$_{1/2}$\,=\,0.90$^{+0.23}_{-0.16}$ ms. This fission activity with
	T$_{1/2}$\,=\,0.9$^{+0.6}_{-0.3}$ ms
	was also observed by Khuyagbaatar et al. \cite{Khuyag20}. The fission activity was attributed to the ground state decay of $^{245}$Md, and the $\alpha$ activity to an isomeric
	state $^{245m}$Md \cite{Ninov96}. 
	Previously known data on $^{245}$Md were not mentioned in \cite{Pore20}.
	For completeness it should be noted that on the basis of detailed spectroscopic investigation of odd-mass mendelevium isotopes performed since then \cite{Hess05}
	the $\alpha$ activity would nowadays rather be attributed to $^{245g}$Md and the fission activity to $^{245m}$Md. It further was shown in \cite{Hess05} that $\alpha$ decay in odd mass
	mendelevium isotopes populates predominantly the 7/2$^{-}$[514] Nilsson level in the einsteinium daughter nuclei which decay into the 7/2$^{+}$[633] Nilsson - level and the 
	9/2$^{+}$ member of the rotational band built up on it. As the 9/2$^{+}$ level decays by highly converted M1 transitions into the 7/2$^{+}$ bandhead, the line at 
	E$_{\alpha}$\,=\,8680\,$\pm$20 keV reported in \cite{Ninov96} is thus certainly the result of energy summing of $\alpha$ particles and conversion electrons.
	
	\section{Comparison of the results for $^{245}$Md reported by Ninov et al. \cite{Ninov96} and Khuyagbaatar et al. \cite{Khuyag20} and for $^{244}$Md reported
		by Pore et al. \cite{Pore20}. }
	\label{intro}
	
	The data published for $^{245}$Md in \cite{Ninov96,Khuyag20} and $^{244}$Md in \cite{Pore20} are presented in fig. 1 and table 1. Data of Pore et al. 
	(P1\,-\,P6) are taken 
	from table 1 in \cite{Pore20}. Data of Khuyagbataar et al. (K1\,-\,K10) are taken from the supplemental
	material of \cite{Khuyag20}. No list of single events was presented by Ninov et al.
	\cite{Ninov96}.
	Data shown here (N1\,-\,N8) are taken from a re-inspection of the logbook of the corresponding SHIP experiment \cite{Hess20}. Only $\alpha$ - $\alpha$ correlations with 
	full energy release of both $\alpha$ particles in the SHIP 'stop - detector' are listed.\\
	
	\begin{figure}
		\resizebox{0.9\textwidth}{!}{%
			\includegraphics{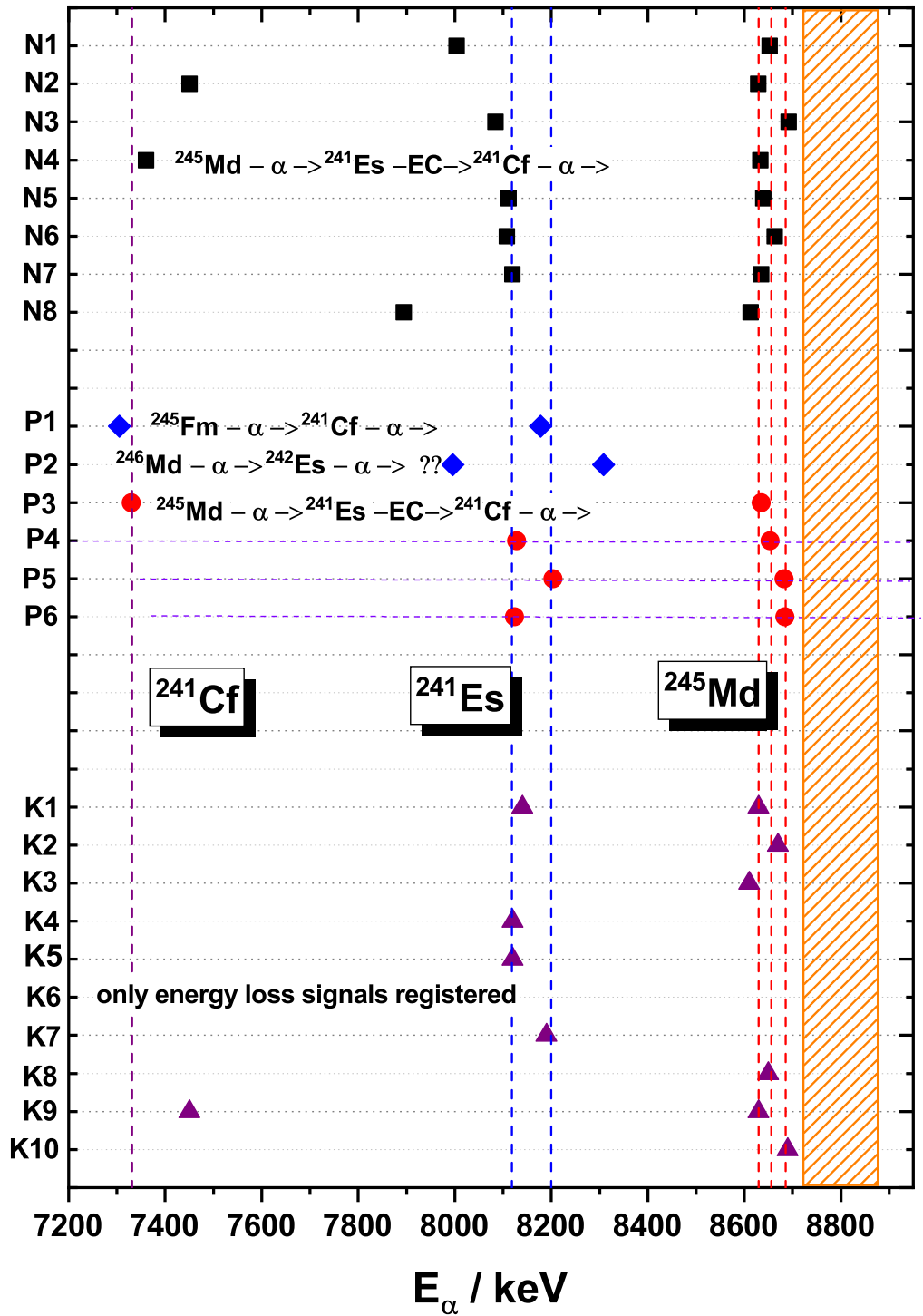}
		}
		\caption{Summary of decays attributed to $^{245}$Md in \cite{Ninov96}(squares) as well as in \cite{Khuyag20}(triangles) together with data  reported by Pore et al. \cite{Pore20} (circles: events attributed to $^{245}$Md by the present authors, diamonds: events attributed to $^{245}$Fm or (tentatively) to $^{246}$Md). The dashed lines are to guide the eyes: the red lines represent the $\alpha$ energies given for $^{245}$Md (8640, 8680 keV) in \cite{Ninov96} and the energy given for $^{244}$Md (8663 keV) in \cite{Pore20}; the blue lines represent the $\alpha$ energy for $^{241}$Es (8113 keV) given in \cite{Ninov96} and the highest daughter
		energy (P5) in \cite{Pore20}; the purple line repesents the literature value of the $\alpha$ energy of $^{241}$Cf (7335 keV) \cite{Fire96}. The orange
		hetched area marks the range of $\alpha$ energies where the events attributed to $^{244}$Md in \cite{Khuyag20} were observed.}
		\label{fig:1}       
	\end{figure}

	\begin{table}
		\caption{Summary of decays attributed to $^{245}$Md in \cite{Ninov96,Khuyag20} and decays reported by Pore et al. \cite{Pore20}.
			Data from Pore et al. are taken from table 1 in \cite{Pore20}; data from Khuyagbaatar et al. are from the supplemental material \cite{Khuyag20}.
			No individual decay data are reported in \cite{Ninov96}; these data are taken from the experiment analysis logbook \cite{Hess20}.}
		\label{tab:3}       
		\footnotesize{
			\begin{tabular}{llllll}
				\hline\noalign{\smallskip}
				\noalign{\smallskip}\hline\noalign{\smallskip}
				Ref. & evt. no. & E$_{\alpha}$(1)/MeV  & $\Delta$t(ER-$\alpha$1)/s & E$_{\alpha}$(2)/MeV  & $\Delta$t($\alpha$1-$\alpha$2)/s \\ 
				\hline\noalign{\smallskip}     
				\cite{Ninov96}  & N1 &  8.652 & 0.0178 & 8.004 &  8.254   \\ 
				\cite{Ninov96}  & N2$^{*}$ &  8.629 & 0.1751 & 7.450 &  88.083   \\ 
				\cite{Ninov96}  & N3 &  8.692 & 0.00164 & 8.084 &  28.406   \\ 
				\cite{Ninov96}  & N4 &  8.633 & 0.1565 & 7.360 &  203.876   \\ 
				\cite{Ninov96}  & N5 &  8.639 & 1.1708 & 8.111 &  7.639   \\ 
				\cite{Ninov96}  & N6 &  8.663 & 0.0843 & 8.108 &  15.763   \\ 
				\cite{Ninov96}  & N7 &  8.635 & 0.2831 & 8.119 &  13.573   \\ 
				\cite{Ninov96}  & N8 &  8.613 & 0.0914 & 7.894 &  335.005   \\ 
				\hline\noalign{\smallskip}
				\cite{Khuyag20}  & K1$^{**}$ &  8.63 & 0.564 & 8.14 &  4.73   \\ 
				\cite{Khuyag20}  & K2$^{**}$ &  8.67 & 0.454 & (1.1) &  0.24   \\ 
				\cite{Khuyag20}  & K3$^{**}$ &  8.61 & 0.423 & (1.3) &  2.86   \\ 
				\cite{Khuyag20}  & K4$^{**}$ &  (1.9) & 0.120 & 8.12 &  6.87   \\ 
				\cite{Khuyag20}  & K5$^{**}$ &  (2.2) & 0.508 & 8.12 &  11.5   \\ 
				\cite{Khuyag20}  & K6$^{**}$ &  (0.9) & 0.131 & 8.09 & 15.1   \\ 
				\cite{Khuyag20}  & K7$^{**}$ &  (0.4) & 1.42 & 8.19 &  2.97   \\ 
				\cite{Khuyag20}  & K8$^{***}$ &  8.65 & 0.693 & (0.26) &  5   \\ 
				\cite{Khuyag20}  & K9$^{***}$ &  8.63 & 0.346 & 7.45 &  20   \\ 
				\cite{Khuyag20}  & K10$^{***}$ & 8.69 & 0.129 &  miss. &  miss.   \\ 
				\hline\noalign{\smallskip}
				\cite{Pore20}  & P1 &  8.178 & 0.60 & 7.305 &  27.34   \\ 
				\cite{Pore20}  & P2 &  8.308 & 9.18 & 7.996 &  14.37   \\ 
				\cite{Pore20}  & P3 &  8.635 & 0.88 & 7.330 &  18.95   \\ 
				\cite{Pore20}  & P4$^{****}$ &  8.653 & 0.13 & 8.128 &  1.20   \\ 
				\cite{Pore20}  & P5 &  8.682 & 0.31 & 8.203 &  10.00   \\
				\cite{Pore20}  & P6 &  8.684 & 1.16 & 8.124 &  7.65   \\ 
				
			\end{tabular}
		}
		\vspace*{0.5cm}  
		\\
		$^{*}$ Both events were registered within the beam on period\\
		$^{**}$ observed at E$^{*}$\,=\,26.2 MeV\\
		$^{***}$ observed at E$^{*}$\,=\,32.7 MeV \\
		$^{****}$ the $\alpha$ - $\alpha$ correlation was followed by a third event of 
		E$_{\alpha}$\,=\,7.086$\pm$25 MeV after $\Delta$t\,=\,75.97 s.
		
	\end{table}

	\begin{figure}
		\resizebox{0.90\textwidth}{!}{%
			\includegraphics{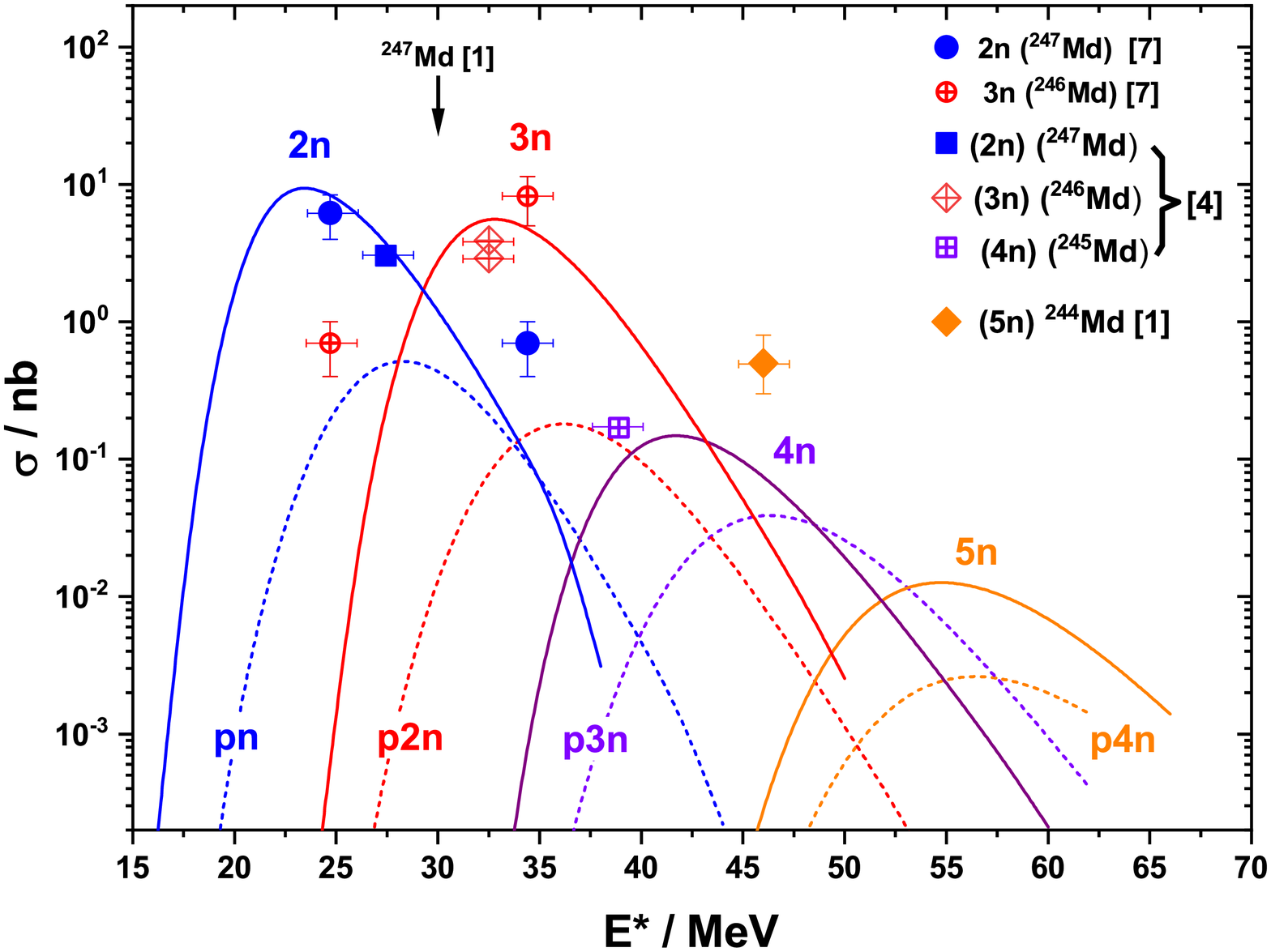}
		}
		\caption{Excitation function for $^{40}$Ar + $^{209}$Bi. The energies refer to production in the center of the target.
		The error bars for the energies refer to the energy loss of $^{40}$Ar ion in the bismuth targets \cite{Ziegler13}.
		Systematic errors in the accelerator energy are typically 0.2$\%$ for the UNILAC accelerator and are neglected. 
		For the data of Pore et al. \cite{Pore20} an energy loss of $\approx$12.5 MeV in the titanium backing foil \cite{Ziegler13} is 
		considered. No systematic error for the accelerator energy is given by Pore et al..
		Lines are the result of HIVAP \cite{Reis92} calculations; full lines represent xn-channels,
		dashed lines represent pxn-channels.
		Points are defined in the figure. The arrow marks the energy reported in \cite{Pore20} for the observation of $^{247}$Md.
	}
		\label{fig:2}       
	\end{figure}
	
	Evidently the chains P3, P4 and P6 agree with the data reported for $^{245}$Md in \cite{Ninov96,Khuyag20}.
	The energy of the daughter in P5 is higher than the values reported for $^{241}$Es in \cite{Ninov96},
	but is in agreement with the daughter energy in K7. This event was attributed to $^{245}$Md in \cite{Khuyag20} as it was registered 
	at  E$^{*}$\,=\,26.2 MeV, where only decays of $^{245}$Md were observed. Concerning the daughter energies
	P4, P5 and P6 can be attributed to the decay $^{245}$Md $^{\alpha}_{\rightarrow}$ $^{241}$Es $^{\alpha}_{\rightarrow}$, 
	while P3 obviously represents the decay  $^{245}$Md $^{\alpha}_{\rightarrow}$ $^{241}$Es $^{EC}_{\rightarrow}$ $^{241}$Cf $^{\alpha}_{\rightarrow}$,
	in accordance with N4 and the known $\alpha$ decay energy of $^{241}$Cf (7.340 MeV \cite{Fire96}).
	P1 fits to the decay sequence $^{245}$Fm $^{\alpha (8.15 MeV)}_{~~~~\rightarrow}$ $^{241}$Cf $^{\alpha (8.34 MeV)}_{~~~~\rightarrow}$ \cite{Fire96},
	with $^{245}$Fm being the product of the p3n - channel. The cross-section ratio $\sigma$(p3n)/$\sigma$(4n)\,$\approx$0.25 may appear unusually high, 
	but it has to be considered that one approaches the proton drip-line, and proton binding energies are already low. The mass evaluation
	of Wang et al. \cite{Wang16} delivers values of, e.g., 1540$\pm$210 keV for $^{247}$Md and 1360$\pm$320 keV for $^{246}$Md, significantly lower 
	than the neutron binding energies of 8250$\pm$330 keV ($^{247}$Md) and 7230$\pm$400 keV ($^{246}$Md).
	And indeed HIVAP calculations \cite{Reis92} deliver even a ratio $\sigma$(p3n)/$\sigma$(4n)\,$\approx$0.5 (see fig.2).
	It should be reminded that recently notable cross - sections for p - evaporation channels have been reported for the
	reaction $^{50}$Ti + $^{209}$Bi \cite{Lopez19,Hessb19}.\\ 
	Less clear is chain P2. The decay sequence $^{246}$Md $^{\alpha}_{\rightarrow}$ $^{242}$Es  $^{\alpha}_{\rightarrow}$,
	for which very broad energy distributions in the range E$_{\alpha}$\,$\approx$\,(8.15-8.75) MeV ($^{246}$Md) and 
	E$_{\alpha}$\,$\approx$\,(7.75-8.05) MeV ($^{242}$Es) were observed (see fig. 5 in \cite{Anta10})
	is a possible candidate.\\
	P4 is terminated by an $\alpha$ event of E$_{\alpha}$\,=\,7.086$\pm$25 MeV, which could be attributed to $^{237}$Bk, the so far unknown $\alpha$ daughter
	of $^{241}$Es. From atomic mass extrapolation \cite{Wang16} one expects an $\alpha$ decay energy of E\,=\,7.376$\pm$0.242 MeV. The lower value could be due to the population of an excited 
	state in $^{233}$Am.
	
	\section{Excitation functions. }
	The reported cross sections for production of $^{244-247}$Md in the reaction $^{209}$Bi($^{40}$Ar,xn)$^{249-x}$Md \cite{Anta10,Pore20,Hess20} are shown
	in fig. 2. In \cite{Ninov96} no cross sections are given. The values given for this experiment are taken from \cite{Hess20}.
	The lines are the result of HIVAP \cite{Reis92} calculations, using fission barriers modified to reproduce the 2n ($^{247}$Md) and 3n ($^{246}$Md) 
	cross sections. Evidently the 4n - cross section from \cite{Hess20} is reproduced quite well. The excitation energy given by Pore et al. \cite{Pore20} appears roughly 
	4 MeV above the expected maximum for the 4n - cross section, and the value is about a factor of six higher, but more than two orders of magnitude higher than the value
	expected for the 5n - channel.\\
	A similiar situation is evident for the 2n - channel. Pore et al. \cite{Pore20} report the observation of $^{247}$Md at a bombarding energy of 200 MeV, which 
	corresponds to an excitation energy E$^{*}$$\approx$30 MeV (arrow in fig. 2), which is about 6 MeV above the expected maximum for the 2n channel,
	but still a notable production cross - section of $\approx$2 nbarn is expected here.\\
	To conclude: comparison with reported cross-sections for xn - channels and HIVAP calculations indicates that the events attributed to $^{244}$Md in \cite{Pore20}
	may rather stem from decay of $^{245}$Md.
	
	\section{Conclusion. }
	The decay data for $^{244}$Md presented by Pore et al. \cite{Pore20} are in disagreement with those published by Khuyagbaatar et al. \cite{Khuyag20}.
	A critical inspection of the decay data of Pore et al. \cite{Pore20} for $^{244}$Md and a comparison with reported decay data for $^{245}$Md rather suggest that
	they have observed $^{245}$Md. An additional argument supporting that interpretation comes from the excitation function for the production of mendelevium isotopes 
	in the reaction $^{40}$Ar + $^{209}$Bi. The excitation energy given for the observation of $^{244}$Md is about 10 MeV lower than the expected maximum for the 5n - channel.
	Bombarding energy and 
	reported production cross section rather hint at the synthesis of $^{245}$Md. 
	


\begin{thebibliography}{00}
		\bibitem{Pore20}
		J.L. Pore et al., Phys. Rev. Lett. {\bf 124}, 252502 (2020).
		%
		\bibitem{Khuyag20}
		J. Khuyagbaatar et al., Phys. Rev. Lett. {\bf 125}, 142504 (2020).
		%
		\bibitem{Ninov96}
		V. Ninov et al., Z. Phys. A {\bf 356}, 11 (1996).
		%
		\bibitem{Hess20}
		F.P. He\ss berger, Analysis Logbook SHIP experiment R165 (1993).
		%
		%
		\bibitem{Hess05}
		F.P. He\ss berger et al., Eur. Phys. J. A {\bf 26}, 233 (2005).
		%
		%
		\bibitem{Fire96}
		R.B. Firestone et al., Table of Isotopes, 8th Edition,
		John Wiley $\&$ Sons, New York (1996).
		%
		%
		\bibitem{Anta10}
		S. Antalic et al., Eur. Phys. J. A {\bf 43}, 35 (2010).
		%
		%
		\bibitem{Wang16}
		M. Wang et al., Chinese Phys. C{\bf 41}, 03003 (2016).
		%
		\bibitem{Reis92}
		W. Reisdorf, M. Sch\"adel, Z. Phys. J. A {\bf 343}, 47 (1992).
		%
		\bibitem{Lopez19}
		A. Lopez-Martens et al., Phys. Lett. B {\bf 795}, 271 (2019).
		%
		\bibitem{Hessb19}
		F.P.He\ss berger, Eur. Phys. A {\bf 55:} 208 (2019).
		%
		\bibitem{Ziegler13}
		J.F. Ziegler, J.P. Biersack, M.D. Ziegler, SRIM-2013.00,
		http://srim.org/ (2013)
		
		
	\end{thebibliography}
\end{document}